\documentclass[%
 reprint,
 amsmath,amssymb,
 aps,
]{revtex4-1}

\usepackage{color}
\usepackage{graphicx}
\usepackage{dcolumn}
\usepackage{bm}

\begin{document}

\preprint{APS/123-QED}

\title{Non-Fermi liquid behavior of electrical resistivity close to the nematic critical point in Fe$_{1-x}$Co$_x$Se and FeSe$_{1-y}$S$_y$}

\author{T. Urata$^1$}
 \thanks{Corresponding author: urata@nuap.nagoya-u.ac.jp}
 \thanks{\\Present address: Department of Crystalline Materials Science, Nagoya University, Nagoya 464-8603, Japan.}
\author{Y. Tanabe$^1$}
 \thanks{Corresponding author: ytanabe@m.tohoku.ac.jp}

\author{K. K. Huynh$^2$}

\author{H. Oguro$^3$}

\author{K. Watanabe$^3$}

\author{K. Tanigaki$^{1, 2}$}
\thanks{Corresponding author: tanigaki@m.tohoku.ac.jp}

\affiliation{$^1$Department of Physics, Graduate School of Science, Tohoku University, Sendai 980-8578, Japan}

\affiliation{$^2$WPI-Advanced Institutes for Materials Research, Tohoku University, Sendai 980-8577, Japan}

\affiliation{$^3$High Field Laboratory for Superconducting Materials, Institute for Materials Research, Tohoku University, Sendai 980-8577, Japan}

\date{\today}

\begin{abstract}
Temperature dependence of resistivity of single crystals of Fe$_{1-x}$Co$_x$Se and FeSe$_{1-y}$S$_y$ is studied in detail under zero and high magnetic field (magnetoresistance), the latter of which enables to monitor the temperature ($T$) evolution of resistivity below the onset of superconducting transition temperature ($T_{\rm c}$).
In FeSe$_{1-y}$S$_y$, $T$-linear dependence of resistivity is prominent in $y$ = 0.160 below 40 K, whereas it changes to a Fermi-liquid(FL)-like $T^2$ one below 10 K in $y$ = 0.212.
These suggest that the quantum critical point (QCP) originating from the electronic nematicity resides around $y$ = 0.160 and the fluctuation in QCP gives rise anomalous $T$-linear dependence in resistivity in a wide $T$ range.
In Fe$_{1-x}$Co$_x$Se, resistivity gradually changes from linear- to quadratic- $T$-dependent one at low temperatures in the range between $x$ = 0.036 and 0.075.
These could be interpreted by scenarios of both the nematic QCP and the crossover in the ground states between the orthorhombic nematic phase and the tetragonal phase.
The anomalies found as $T$-linear resistivity are discussed in terms of orbital and spin fluctuation arising from the nematic QCP.
\end{abstract}

\maketitle

%
The iron based superconductors (FeSCs) provide the new paradigm for the physics in the high temperature superconductivity where the orbital, the spin, and the lattice degree of freedom are considered to inextricably participate to the formation of the Cooper pair.
Most parent compounds of FeSCs show the $C_4$ symmetry breaking (nematic) transition accompanied by the stripe type antiferromagnetic (AFM) order \cite{Greene,Fernandes,Hosono_Kuroki}.
The superconductivity often emerges when the order is suppressed under pressure or/and the chemical substitutions.
The important experimental fact is that a superconducting dome takes a maximum at the quantum critical point (QCP) of the AFM order in their electronic phase diagrams.
Consequently, AFM quantum fluctuation has been considered to play important role for the formation of Cooper pairs.

From such aspects, a considerable number of researches have been made on the BaFe$_2$(As$_{1-x}$P$_x$)$_2$ systems.
Non-Fermi liquid (nFL) behaviors have been identified in several physical properties e.g. penetration depth and resistivity slope in the optimally doped region \cite{Kasahara_prb,Nakai,Hashimoto1554,Walmsley,Analytis_RT}.
This indicates that a strong AFM fluctuation is present near the QCP and substantially contributes to the formation of the Cooper pair.
Nevertheless, the relationship between the nematic quantum fluctuation and the superconductivity has been an open question, because unique characters smear out when the AFM ordering simultaneously emergent.
In order to answer to this question, FeSe may be the most appropriate system where the orbital order takes place at around structural transition temperature ($T_{\rm s} \approx 90$ K) without the AFM order \cite{Hsu_FeSe,McQueen}, hence non-magnetic nematic state is realised \cite{Huynh,Nakayama,Bohmer_NMR,Baek,Suzuki}.
Recent experimental progresses in single crystalline FeSe$_{1-y}$S$_y$ have revealed that the nematic fluctuation diverges towards the nematic QCP where the nematic quantum transition takes place \cite{Hosoi}.
It is required to investigate physical phenomena evoked from the nematic fluctuation in FeSCs.
In addition, the relationship between the non-magnetic nematic QCP and the superconductivity is an urgent subject to elucidate.
In this report, we study cobalt and sulfur substitution effects on electrical resistivity in FeSe single crystals.
If low energy excitations are governed by certain quantum fluctuations, the temperature dependence of resistivity is expected to show anomalous nFL behaviors \cite{Ueda,Hertz,Millis}.
We try to find such evidence from the viewpoint of the evolution of resistivity as a function of $T$.
In order to detect the experimental facts, the normal-state resistivity at temperatures lower than the onset of superconducting transition temperature ($T_{\rm c}$) and the magnetoresistance (MR) are measured at various temperatures and their values in the normal-states were extrapolated to zero field.
Almost $T$-linear dependence of resistivity is found at around the nematic QCP in FeSe$_{1-y}$S$_y$, whereas deviation was observed in Fe$_{1-x}$Co$_x$Se.
In the overdoped regime where the orbital order disappears, a quadratic temperature dependence of resistivity emerges towards the low-$T$ limit, being indicative of the crossover from the nFL to the FL at low temperatures.
The origin of the changes observed in resistivity is discussed in terms of orbital and spin fluctuations.

\section{Experiments}
High quality single crystals of Fe$_{1-x}$Co$_x$Se ($0 \leq x \leq 0.08$) and FeSe$_{1-y}$S$_y$ ($0 \leq y \leq 0.21$) were grown by a molten salt flux method \cite{Bohmer_synth,Huynh,Nakayama,Urata_Co}.
As precursors, polycrystalline samples were synthesized by a solid state reaction \cite{Mizuguchi}.
The quality of the single crystals was examined by the (0 0 $l$) reflection of X-ray diffraction (XRD) and energy dispersive X-ray spectroscopy (EDS).
The EDS spectra were taken at different ten points for each sample and the averaged molar ratio among iron, selenium, and cobalt ($x$) or sulfur ($y$) was calculated.
The errors of compositions were estimated using the standard deviations from the averaged values.
Temperature dependence of the $\rho$, MR and Hall resistivity were measured by the standard four probe method.
Magnetic fields ($B$) were varied with $|B|\leq 18$ T paralleled to the c-axis.
$T_{\rm c}$s are defined at the end point of the superconducting transition where $\rho$ is approximately less than 1.0$\times 10^{-8}$ $\Omega$cm.
Note that MR and Hall resistivity are averaged or subtracted between positive and negative $B$ respectively to cancel unnecessary contributions from antisymmetric or symmetric components owing to the misalignment of electrodes.

\section{Results and discussion}

\begin{figure}
\includegraphics[bb=0 0 636 555,width=1.0\linewidth]{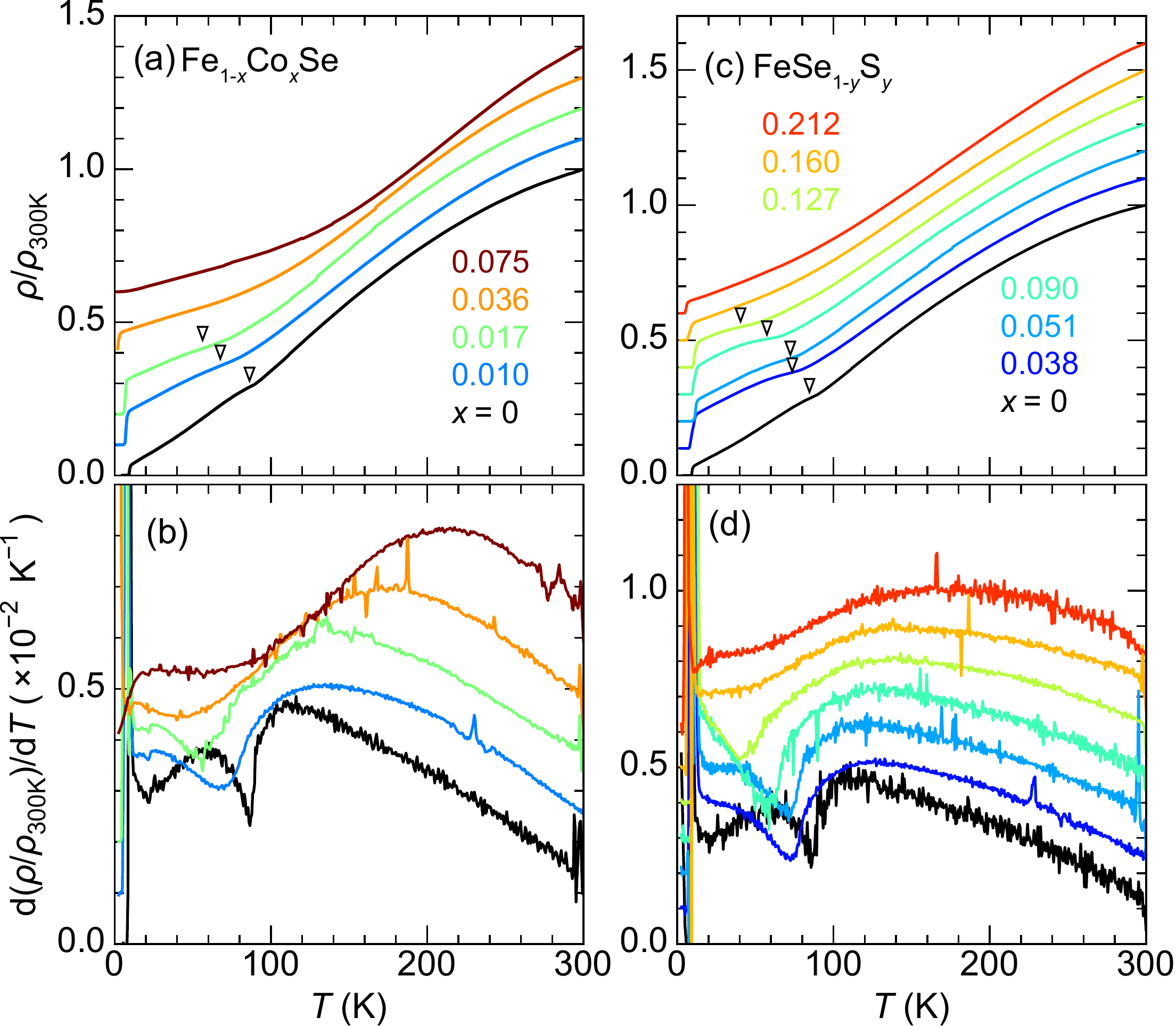}
\caption{(a,c) The normalized temperature dependence of resistivity ($\rho/\rho_{\rm 300K}$) for (a) Fe$_{1-x}$Co$_x$Se (0 $\leq$ $x$ $\leq$ 0.075) and (c) FeSe$_{1-y}$S$_y$ (0 $\leq$ $y$ $\leq$ 0.212).
Note that each curve is shifted by 0.1 for clarity.
The down open triangles indicate the kink point in the resistivity curve.
They are determined from the peak position of the 1st derivative.
(b,d) The temperature dependence of 1st derivative of $\rho/\rho_{\rm 300K}$.}
\label{res}
\end{figure}

\begin{figure*}
\includegraphics[bb=0 0 844 325,width=1.0\linewidth]{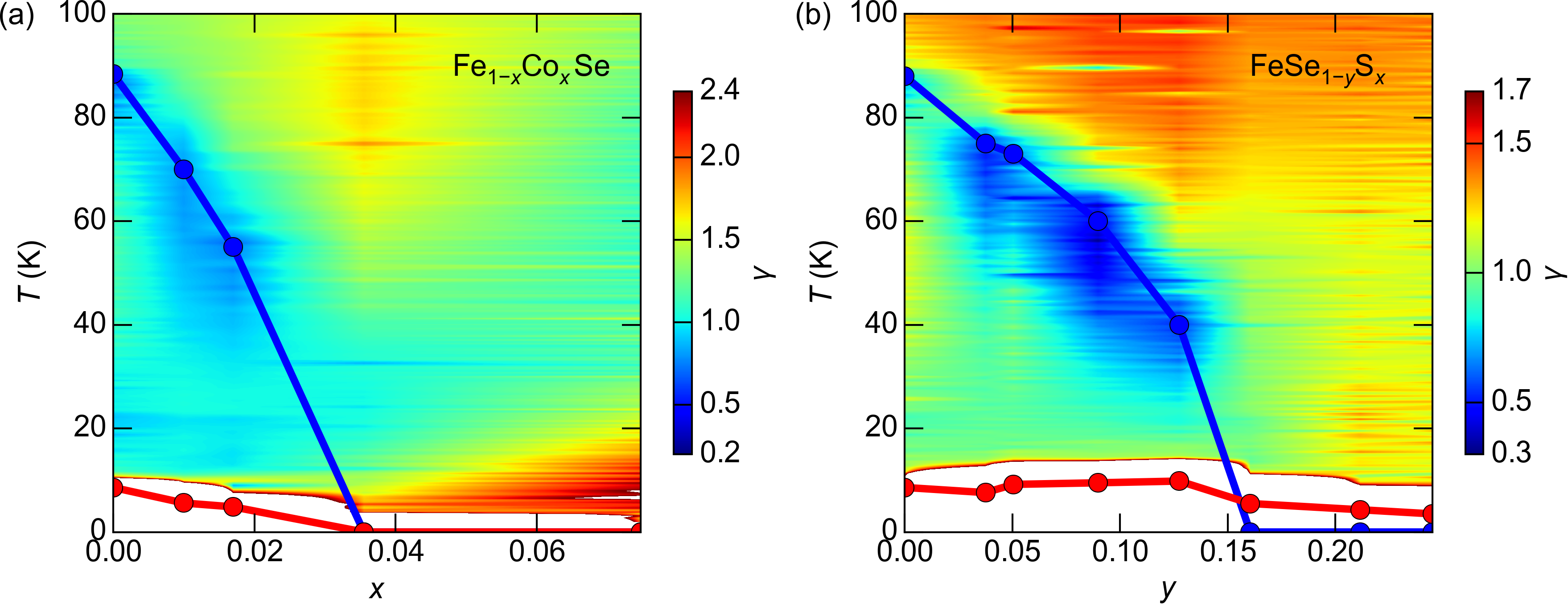}
\caption{The electronic phase diagram of (a) Fe$_{1-x}$Co$_x$Se (0 $\leq$ $x$ $\leq$ 0.075) and (b) FeSe$_{1-y}$S$_y$ (0 $\leq$ $y$ $\leq$ 0.212).
The blue and red circles connected by lines are the structural transition temperatures ($T_{\rm s}$) and superconducting transition temperatures ($T_{\rm c}$) both estimated from resistivity measurements, respectively.
The temperature ($T$) exponents of resistivity curves ($\gamma$) are shown as color maps.
The data shown in intermediate concentrations are plotted by interpolating linearly.
The blank region indicates $\gamma$ takes value beyond the color scale due to the superconducting transition.
}
\label{contour}
\end{figure*}

Fig. \ref{res} (a, c) shows temperature dependences of the normalized resistivity at 300 K ($\rho/\rho_{\rm 300K}$) for Fe$_{1-x}$Co$_x$Se and FeSe$_{1-y}$S$_y$, respectively.
The superconducting transition was observed in Fe$_{1-x}$Co$_x$Se ($0 \leq x \leq 0.017$) and FeSe$_{1-y}$S$_y$ ($0 \leq y \leq 0.212$).
In FeSe$_{1-y}$S$_y$, $T_{\rm c}$ shows the maximum at around $x$ = 0.127 and gradually decreases with an increase in $y$, which is in sharp contrast with that of Fe$_{1-x}$Co$_x$Se, where $T_{\rm c}$ monotonically decreases with an increase in $x$.
Above $T_{\rm c}$, we find a kink in the resistivity curve for samples with various compositions.
In order to show this more clearly, the first derivatives of $\rho/\rho_{\rm 300K}$ are depicted in Fig.\ref{res} (b, d).
The values of d($\rho/\rho_{\rm 300K}$)/d$T$ show sharp changes in derivative for Fe$_{1-x}$Co$_x$Se ($x \leq 0.017$) and FeSe$_{1-y}$S$_y$ ($y\leq 0.127$).
We name these temperatures as $T_{\rm kink}$ (shown as the triangles in the top figures of Fig.\ref{res}).
Looking back at the previous reports, $T_{\rm kink}$ most presumably corresponds to $T_{\rm s}$ \cite{McQueen}.
Hence, the structural transition seems to be suppressed by both cobalt and sulfur doping and disappears in the stoichiometry around $x \approx 0.036$ and $y \approx 0.160$, respectively.

As is mentioned in recent reports, FeSe family may possess non-magnetic nematic QCP in the phase diagram \cite{Huynh,Nakayama,Bohmer_NMR,Baek,Suzuki,Hosoi}.
In the vicinity of QCP, physical properties are frequently described beyond the FL theory.
In the normal FL, the current is carried by quasiparticles and the resistivity is proportional to $T^2$ at low temperatures \cite{Landau_FL}.
On the other hand, if the system is influenced by quantum fluctuations, differently exotic $T$ dependence of resistivity emerges.
In order to see the nematic quantum fluctuations in FeSe, the resistivity curves are analyzed in detail by focusing on around the nematic QCP in the phase diagram.

We carefully examined temperature evolution of the resistivity as follows.
The temperature dependence of resistivity ($\rho$) is assumed in the form of $\rho = aT^\gamma + \rho_0$, where $\rho_0$ denotes the residual resistivity and $a$ is a coefficient.
In this assumption, the temperature exponent of resistivity ($\gamma$) can be written as, $\gamma = {\rm dln}(\rho-\rho_0)/{\rm dln}T$.
We evaluated $\gamma$ as a function of temperature by employing this formula.
In Fig. \ref{contour}, the temperature and chemical concentration dependences of the exponent are shown as color maps.
From these phase diagrams, the nematic transition is greatly suppressed around at $x$ $\sim$ 0.036 and $y$ $\sim$ 0.160 in Fe$_{1-x}$Co$_x$Se and FeSe$_{1-y}$S$_y$, respectively.
At around these doping regimes, the onset of $T_{\rm c}$ still remains in both samples and hides the normal-state resistivity behavior at low temperatures.
To clarify the resistivity in the normal state to be viewed from the superconducting transition, we performed transverse MR measurements.

\begin{figure}
\includegraphics[bb=0 0 406 958,width=0.85\linewidth]{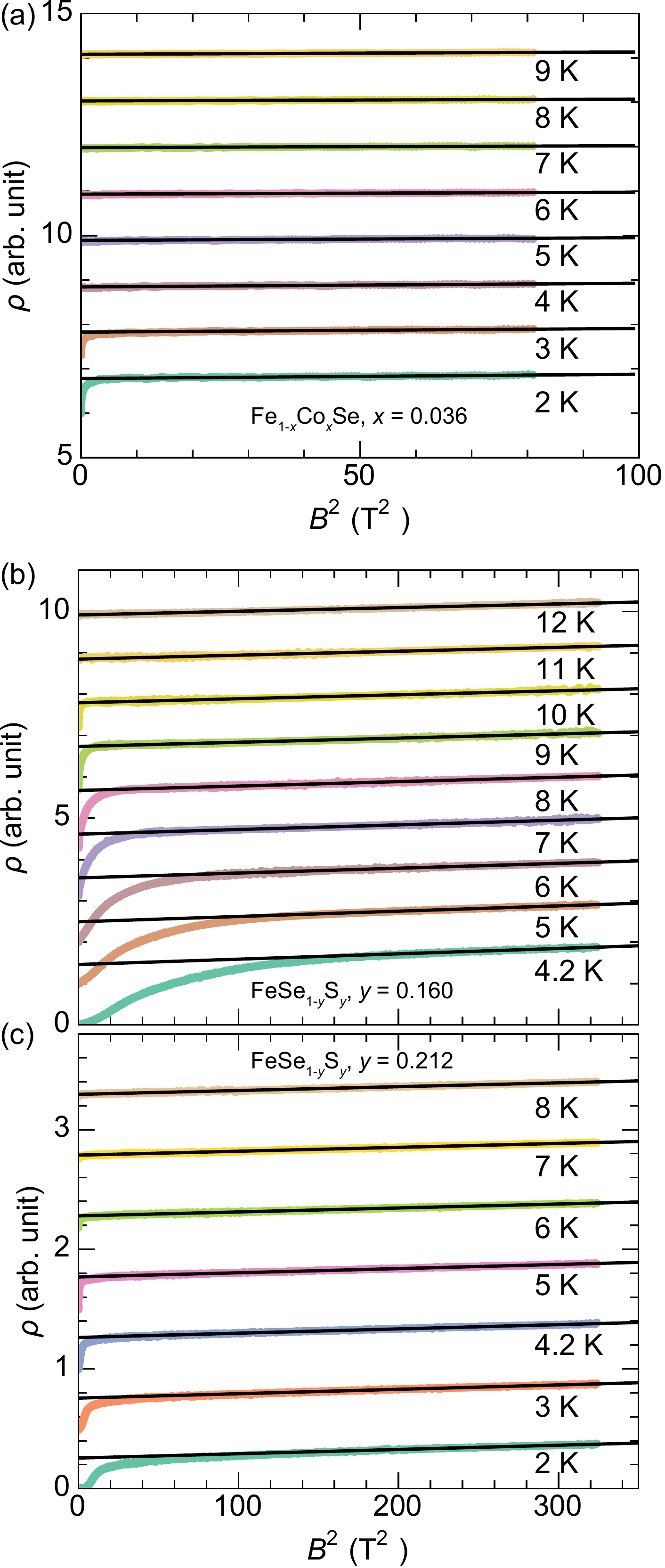}
\caption{The squared magnetic field ($B^2$) dependence of resistance ($\rho$) for (a) Fe$_{1-x}$Co$_x$Se ($x = 0.036$) and (b, c) FeSe$_{1-y}$S$_y$ ($y = 0.160, 0.212$, respectively) at various temperature.
Note that each curve is shifted by arbitrary values for clarity.
The $B^2$-linear lines indicate fitting on high $B$ region.
The zero field intercepts of this fitting lines defined resistivity of normal state at corresponding temperature.
}
\label{MR}
\end{figure}

Fig. \ref{MR} shows the squared magnetic field ($B^2$) dependence of MR at various temperatures for Fe$_{1-x}$Co$_x$Se ($x = 0.036$) and FeSe$_{1-y}$S$_y$ ($y = 0.160, 0.212$).
The magnetic field was scanned up to $\pm$9 T for Fe$_{1-x}$Co$_x$Se ($x = 0.036$) and $\pm$18 T for FeSe$_{1-y}$S$_y$ ($y = 0.160, 0.212$).
Above the $T_{\rm c}$, MR is proportional to $B^2$.
Below the $T_{\rm c}$, MR is also proportional to $B^2$ above upper critical magnetic field $B_{\rm c2}$.
This implies that the family of the compounds are compensated semimetals and low B approximations are still valid even under $B > B_{c2}$.
Actually, the highly compensated semimetallic band structure of FeSe has been reported above \cite{Nakayama} and under $T_{\rm s}$ \cite{Huynh,Terashima,Kasahara_PNAS}.
The isovalent doping FeSe$_{1-y}$S$_y$ may be similar to FeSe with various $y$.
As for Fe$_{1-x}$Co$_x$Se, however, the electron doping occurs \cite{Urata_Co} and carrier compensation may not be preserved.
Consequently, we can imagine that the carrier mobility of Fe$_{1-x}$Co$_x$Se is greatly suppressed and hence the saturation in MR will not be observed in this field regime.

\begin{figure}
\includegraphics[bb=0 0 360 594,width=1.0\linewidth]{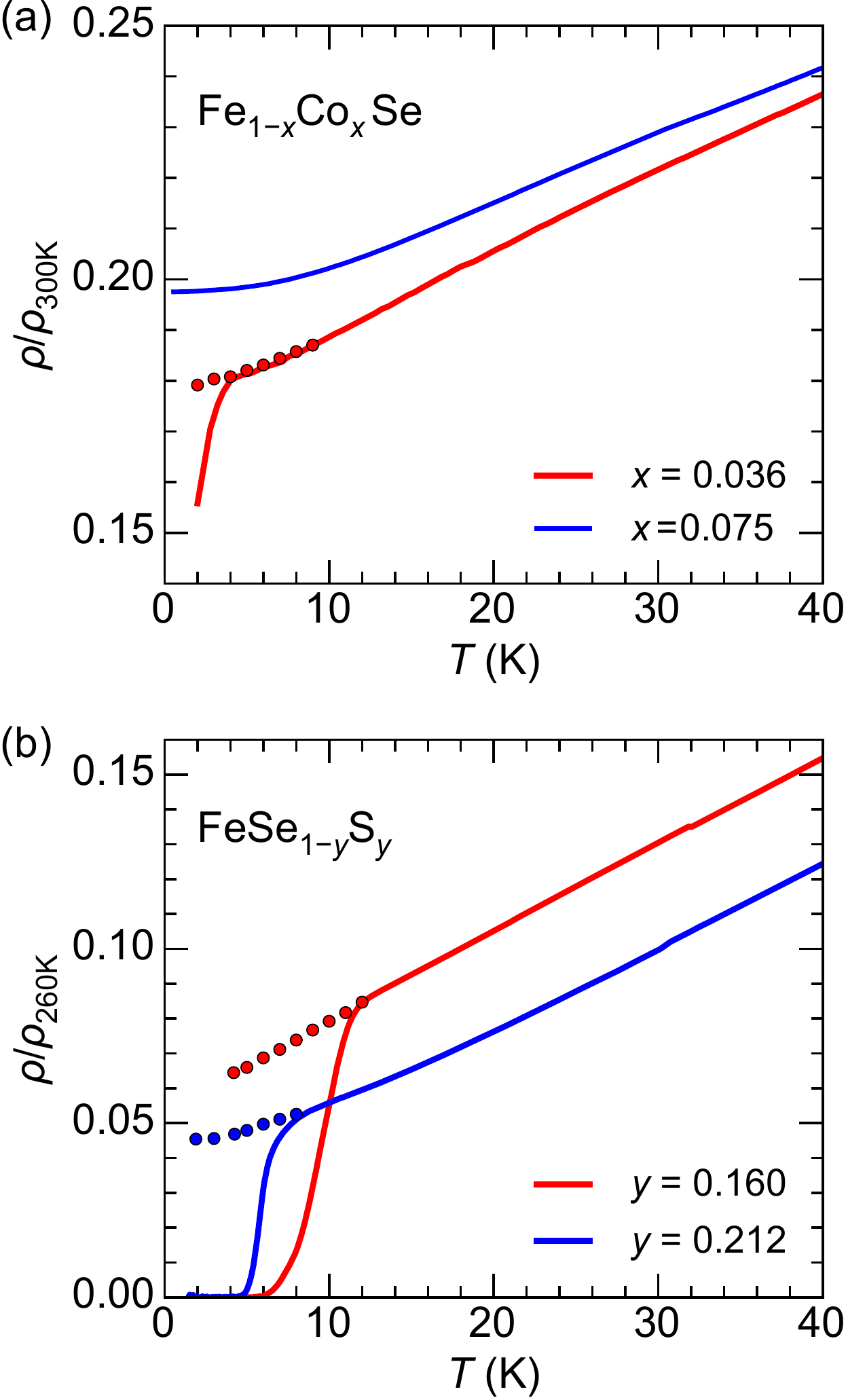}
\caption{The normalized temperature dependence of resistivity ($\rho/\rho_{\rm 300 K}$ and $\rho/\rho_{\rm 260 K}$) for (a) Fe$_{1-x}$Co$_x$Se ($x \leq 0.036, 0.075$) and (b) FeSe$_{1-y}$S$_y$ ($y =  0.160, 0.212$).
The circles indicate the data which are obtained by magnetoresistance measurements.}
\label{res_low}
\end{figure}

The results of $B^2$-linear fitting in the normal-state MR are shown in Fig. \ref{MR}.
The intercepts were estimated by extrapolating the lines to the zero field.
They are plotted with resistivity curves in Fig. \ref{res_low} as circles and are found to be connected smoothly.
We define them as normal-state resistivity at each corresponding temperature.

\begin{figure*}
\includegraphics[bb=0 0 827 355,width=1.0\linewidth]{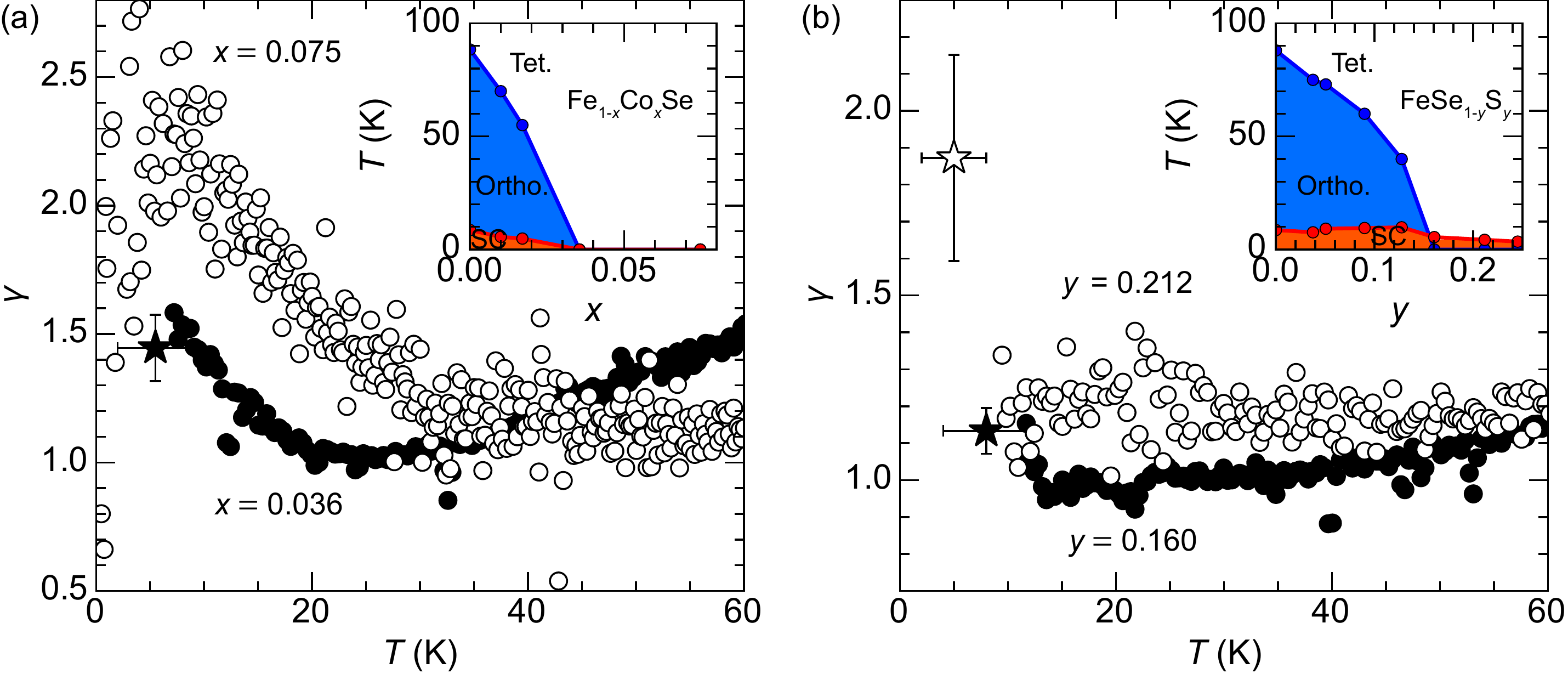}
\caption{The temperature ($T$) dependence of the exponent ($\gamma$) for (a) Fe$_{1-x}$Co$_x$Se ($x = 0.036, 0.075$) and (b) FeSe$_{1-y}$S$_y$ ($y =  0.160, 0.212$).
Note that $\gamma$ are derived from resistivity ($\rho$) $T$ dependences by assuming the function of $\rho = aT^\gamma + \rho_0$, where a is a certain coefficient and $\rho_0$ is the residual resistivity.
Stars denote the data which are obtained from different method (see main text).
Insert figures are phase diagrams drawn by using resistivity data.
}
\label{exp}
\end{figure*}

For Fe$_{1-x}$Co$_x$Se, quadratic evolution of resistivity as a function of $T$ is observed below 10 K in $x$ = 0.036 and this is emphasized in $x$ = 0.075.
For FeSe$_{1-y}$S$_y$, the $\rho$ curves are almost $T$-linear in $y$ = 0.160 below 40 K.
This linear behavior crossovered to the quadratic curve at low temperatures for $y$ = 0.212.
For more quantitative analyses, we again evaluated the temperature exponents by employing the data at low temperatures of these samples.
Nevertheless, at low temperatures, because the intervals of data points are large, the temperature derivatives employed in the above analyses, were not applicable.
Hence, in order to obtain the exponent $\gamma$, we fitted the discrete data plots directly by $\rho = aT^\gamma + \rho_0$.
It is noted that unavoidable errors in the analyses should be taken into account resulting from the larger fitting window for the analyses.
In Fig. \ref{exp}, the stars are the data obtained by the latter method.

Fig. \ref{exp} shows the temperature dependence of $\gamma$ for Fe$_{1-x}$Co$_x$Se ($x = 0.036, 0.075$) and FeSe$_{1-y}$S$_y$ ($y =  0.160, 0.212$).
For FeSe$_{1-y}$S$_y$, $\gamma \approx 1$ is confirmed below 40 K, whereas it deviated to increase toward $\approx 2$ below 10 K in $x$ = 0.212.
The situation is reminiscent of the AFM QCP behavior observed in BaFe$_2$(As$_{1-x}$P$_x$)$_2$ \cite{Kasahara_prb,Nakai,Analytis_RT}, implying the scenario of nematic QCP and fluctuation. 
For Fe$_{1-x}$Co$_x$Se, $T$-linear dependence can be seen from 20 to 30 K for $x = 0.036$ but it increased below 20 K.
In $x$ = 0.075, $\gamma$ gradually increased below 40 K and saturated at around $\sim$ 2 below 10 K.
If the $T$-linear dependence could arise from nematic fluctuation around nematic QCP, $x = 0.036$ would be considered to be higher than the value expected from the exact nematic QCP to some extent.
Consequently, within this framework of interpretation, it appears that the deviation of $\gamma$ from $\approx 1$ in Fe$_{1-x}$Co$_x$Se ($x = 0.036, 0.075$) at low temperatures originates from the crossover between nFL and FL.
In order to understand the observed temperature dependence, another possible scenario would be that the orthorhombic nematic phase gradually crossovers to the tetragonal phase at around $x = 0.036$, where two different features may coexist due to the mixing of nFL and FL states.

Based on the experimental data, we discuss the origin of the anomalous $T$ dependence of $\rho$ at low-$T$ limit in FeSe$_{1-y}$S$_y$.
The $T$-linear resistivity is prominent around the nematic QCP in $y$ = 0.160, whereas it changes to the FL-like one at low temperatures in the overdoped region ($y$ = 0.212).
Therefore, the $T$-linear dependence can be regarded as the consequence of the nematic fluctuations instead of the conventional multi band effects \cite{Albenque}.
Indeed, the change in temperature dependence of $\rho$ in the phase diagram can well be interpreted in terms of the nematic QCP and its fluctuation.
One possible scenario of the nematic order is the orbital order induced by spin fluctuation \cite{Yamakawa_PRX}.
In this case, the spin and orbital fluctuation simultaneously enhanced at the nematic QCP and the nFL resistivity curve would be evoked.
However, no theoretical studies have been made for $\rho - T$ behavior under strong orbital fluctuations so far.
Another scenario is the spin nematic state also induced by spin fluctuation \cite{Fernandes_prl,Wang_natphys,Rong}.
The $\rho$ is known to be $T$-linear under strong two dimensional AFM fluctuations \cite{Ueda,Hertz,Millis}.
This is most likely consistent presently with our experimental results, whereas the spin fluctuation in FeSe has been reported to be small and only develop strongly below the nematic transition temperature \cite{Bohmer_NMR,Baek,Shamoto}.
In order to clarify this point, other experiments to detect the spin fluctuations for chemically doped samples will be important.

\section{Conclusion}
We measured resistivity ($\rho$) temperature dependences on single crystals of Fe$_{1-x}$Co$_x$Se ($0 \leq x \leq 0.08$) and FeSe$_{1-y}$S$_y$ ($0 \leq y \leq 0.21$).
Magnetoresistance (MR) and normal $\rho$ below the onset of the superconducting transition temperature were carefully analyzed.
In FeSe$_{1-y}$S$_y$, $T$-linear $\rho$ was prominent in $y$ = 0.160 below 40 K, whereas it changed to a FL-like one below 10 K in $y$ = 0.212.
These indicate that the nematic QCP resides around $y$ = 0.160 and the quantum critical fluctuations give rise to anomalous $T$-linear $\rho$ in a wide temperature range.
In Fe$_{1-x}$Co$_x$Se, $\rho$ gradually changed from the $T$-linear to the quadratic one at low temperatures in the $x$ range between $x$ = 0.036 and $x$ = 0.075.
These behaviors could be interpreted by the scenarios of both the nematic QCP and the crossover between orthorhombic nematic and tetragonal phases.
In FeSe, both orbital order and spin nematic state have been discussed as primary state under the nematic transition.
Since the $T$-linear $\rho$ around the nematic QCP in FeSe$_{1-y}$S$_y$ may be the consequence of the robust quantum critical fluctuations, its further theoretical interpretation would shed light on the mechanism of the electronic nematicity in FeSe.

\section{Acknowledgements}
This work was performed at High Field Laboratory for Superconducting Materials, Institute for Materials Research, Tohoku University (Project No. 15H0203).
Authors are grateful to K. Nakayama, Y. Yamakawa, and H. Kontani for fruitful discussion.
One of the authors (T.U.) was supported by the Research Fellowship of Japan Society for the Promotion of Science.

\bibliography{FeSeS}

\end{document}